\documentclass[12pt,aps,prb,preprint]{revtex4}   

\usepackage{amsmath}    
\usepackage{graphicx}   

\begin{document}

\title{Cosmic strings  in a model of non-relativistic gravity}
\author{Davood Momeni}
 \affiliation{Department of physics,Faculty of basic
sciences,Tarbiat Moa'llem university,Tehran,IRAN}
 \email{d.momeni@yahoo.com}   
\date{\today}

\begin{abstract}

 Ho$\check{\textbf{r}}$ava proposed a non-relativistic renormalizable theory of
gravitation, which is reduced to  general relativity (GR) in large
distances (infra-red regime (IR)). It is believed that this theory
is an ultra-violet (UV) completion for the classical theory of
gravitation. In this paper, after a brief review of some fundamental
features of this theory, we investigate it for a static cylindrical
symmetric solution which describes \emph{Cosmic string} as a special
case. We have also investigated some possible solutions, and have
seen that how the classical GR field equations are modified for
generic potential $V (g)$. In one case there is an algebraic
constraint on the values of three coupling constants. Finally as a
pioneering work we deduce the most general \emph{cosmic string} in
this theory. We explicitly show that how the \emph{coupling
constants } distort the mass parameter of \emph{cosmic string}. We
deduce an explicit function for mass per unit length of the
space-time as a function of the \emph{coupling constants }. We
compare this function with another  which Aryal et al \cite{58} have
found in GR. Also we calculate the self-force on a massive particle
near Ho$\check{\textbf{r}}$ava-Lifshitz straight string and we give
a typical order for the \emph{coupling constants } $g_{9}$. This
order of magnitude proposes a cosmological test for validity of this
theory.
\end{abstract} \maketitle
\section{Introduction}
In Jaunary 2009, a power-counting renormalizable UV complete theory
of gravity was proposed by Ho$\check{\textbf{r}}$ava
\cite{horava1,horava2,horava3}. Quantum gravity models based on an
"anisotropic scaling" of the space and time dimensions have recently
attracted significant attention\cite{pal,visser}. In particular,
Ho$\check{\textbf{r}}$ava -Lifschitz point gravity \cite{horava1}
might be has desirable features, but in its original incarnation one
is forced to accept a non-zero cosmological constant of the wrong
sign to be compatible with observation\cite{Nastase}. There are four
different versions of this theory :with(or without) projectability
condition and with(or without) detailed balance. In first look it
seems that this non relativistic model for quantum gravity has a
well defined IR limit and it reduces to GR. But as it was first
indicated by Mukohyama\cite{36} ,HL theory mimics GR plus Dark
matter(a pressure less fluid).This theory has a scale invariant
power spectrum which describes inflation. This theory is strongly
coupled and must be modified for escaping  from an unphysical extra
mode.This time this theory has been improved. This theory is
renormalizable in the sense that the effective coupling constant in
the UV is dimensionless. Cosmology in Ho$\check{\textbf{r}}$ava
theory has been studied by several authors
\cite{Kiritsis,Takahashi,Calcagni}. Homogeneous vacuum solutions in
this theory were got in \cite{10}. The cosmological evolution in
Ho$\check{\textbf{r}}$ava gravity with scalar-field was intensively
studied, and the matter bounce scenario in Ho$\check{\textbf{r}}$ava
theory was investigated \cite{11}.\\
Ho$\check{\textbf{r}}$ava theory has at least two important
properties. The first one is it's UV renormalizability, While the
second one is most interesting in cosmology. The fact that the speed
of light diverges in the UV implies that exponential inflation is
not necessary for solving the horizon problem. Moreover, the short
distance structure of perturbations in Ho$\check{\textbf{r}}$ava-
Lifshitz theory is different from standard inflation in GR.
Especially, in UV limit, the scalar field perturbation is
essentially scale-invariant and it is insensitive to the expansion
rate of the universe, as it has been addressed in \cite{Nastase}.In
Ho$\check{\textbf{r}}$ava theory  time and space are treated in an
unequal footing, with four-dimensional general coordinations
invariance emerges as an accidental symmetry in large distance. In
the present form of Ho$\check{\textbf{r}}$ava-Lifshitz cosmology,
one combines the aforementioned modified gravitational background
with a scalar field which reproduces (dark) matter. Doing so we
obtain a dark-matter universe, with the appearance of a cosmological
constant and an effective "dark radiation" term. Although these
terms are interesting cosmological artifacts of the novel features
of Ho$\check{\textbf{r}}$ava-Lifshitz gravitational background, they
restricted the possible scenarios of
Ho$\check{\textbf{r}}$ava-Lifshitz cosmology. Formulating
Ho$\check{\textbf{r}}$ava-Lifshitz cosmology in a way that an
effective dark energy, with a varying equation-of-state parameter,
will emerge is discussed  by Saridakis in \cite{12}. Calcagni found
vacuum solutions and argue that bouncing solutions exist and avoid
the big bang singularity\cite{Calcagni}. \footnote{Solutions with
Euclidean signature are asymptotically de Sitter and in qualitative
agreement with the CDT scenario. On the other hand, inhomogeneous
scalar perturbations against  the classical background, generated by
quantum fluctuations of an inflationary Lifshitz field, are unable
to yield
a scale-invariant spectrum[9]}\\
 The general renormalizable actions
for the scalar field and gauge field are proposed in \cite{13}. They
provide a possible explanation for the time delays in Gamma-Ray
bursts due to the modification of the dispersion relation . Also it
has been shown that the Ho$\check{\textbf{r}}$ava theory for the
completion of General Relativity at UV scales can be interpreted as
a gauge fixed Tensor-Vector theory, and it can be extended to an
invariant theory under the full group of four-dimensional
diffeomorphisms\cite{14}. Charmousis et al  showed that
Ho$\check{\textbf{r}}$ava gravity suffers from strong coupling
problems, with and without detailed balance, and therefore it is
unable to r/heproduce General Relativity in the IR \cite{15}.  Myung
and Kim studied Ho$\check{\textbf{r}}$ava-Lifshitz black hole
solutions and its thermodynamic properties\cite{16}. Mukohyama
presented a simple scenario to generate almost scale-invariant,
super-horizon curvature perturbations\cite{17}. Also  Mukohyama and
et al pointed out that the radiation energy density in the UV epoch
is proportional to $a^{-6}$ and, thus,it decays faster than where in
the IR epoch or in relativistic theories. This leads to intriguing
cosmological consequences such as enhancement of baryon asymmetry
and stochastic gravity waves. They might also discussed current
observational constrains on the dispersion relation\cite{18}.
Topological (charged) black holes Ho$\check{\textbf{r}}$ava-Lifshitz
theory is discussed in \cite{19}. There is a few exact solutions in
Ho$\check{\textbf{r}}$ava theory and it is a considerable problem to
investigate our familiar GR objects in the context of this new
theory. Some exact solutions may be find it \cite{Exact}. Azeyanagi
et al present type IIB super gravity solutions which are expected to
be dual in comparison with certain Lifshitz-like fixed points with
anisotropic scale invariance\cite{44}. Mann found a class of black
hole solutions to a (3+1) dimensional theory gravity coupled to
abelian gauge fields with negative cosmological constant that has
been proposed as a dual theory to a Lifshitz theory describing
critical phenomena in (2+1) dimensions\cite{45}. Ohta and
collaborations discovered new solutions and discussed their
properties\cite{63}.
 Orlando and Reffert studied the
renormalization properties of HL gravity beyond power counting
arguments\cite{60}. In fact, their results confirm its
renormalizability by certain conditions. They make use of the fact
that (super) HL gravity can be taken to the stochastic quantization
of topologically massive gravity. This argument relies on the
renormalizability of the latter, which thought is even not strictly
proven and it is thought to be hold \cite{61}.
Other readable and momentous papers listed in \cite{62}.

Wormhole solutions to Ho$\check{\textbf{r}}$ava theory in   vacuum
are discussed in\cite{31}.The black hole and cosmological solutions
for \emph{arbitrary} cosmological constant was obtained \cite{32}.
One of the best works on thermodynamics of Ho$\check{\textbf{r}}$ava
space times is the paper of Wang and  Wu .They studied
thermodynamics of cosmological models in the
Ho$\check{\textbf{r}}$ava-Lifshitz theory of gravity, and
systematically investigated that the evolution of the universe
filled with a perfect fluid that has the equation of state $p =
w\rho$, where p and $\rho$ denote, respectively, the pressure and
energy density of the fluid, and w is an arbitrary real
constant\cite{33}. Brane cosmology in the
Ho$\check{\textbf{r}}$ava-Witten heterotic M-theory discussed by
 Wu, Gong and  Wang in \cite{43}.
 Too Minamitsuji classified
the cosmological evolutions\cite{34}. The timelike geodesic motion
in the Ho$\check{\textbf{r}}$ava-Lifshitz spacetime studied by Chen
and Wang\cite{35}. Dynamics of a component which behaves like
pressureless dust emerges as an \emph{integration constant} of
dynamical equations investigated by Mukohyama \cite{36}. Saridakis
noted that Ho$\check{\textbf{r}}$ava-Lifshitz cosmology with an
additional scalar field leads to an effective dark energy
sector\cite{37}.The properties of strong field gravitational lensing
in the deformed Ho$\check{\textbf{r}}$ava-Lifshitz black hole
studied by Chen and Jing\cite{38}. Too  Yamamoto et.al studied the
spectral tilt of primordial perturbations in
Ho$\check{\textbf{r}}$ava-Lifshitz
cosmology \cite{46}.\\
But later, Blas et.al\cite{64} listed inconsistencies of the
Ho\v{r}ava-Lifshitz gravity as a complete description of Quantum
gravity. They addressed  \emph{the consistency of Ho\v{r}ava's
proposal for the theory of quantum gravity from the low-energy
perspective}. A peculiarity of the new mode is that \emph{it
satisfies an equation of motion that is of first order in time
derivatives}. In linear level this extra mode manifests only around
spatially inhomogeneous and time-dependent backgrounds. They found
two serious problems associated with this mode. First,\emph{the mode
develops very fast exponential instabilities at short distances}.
Second,\emph{ it becomes strongly coupled at an extremely low cutoff
scale}. They also discussed the \emph{projectable} version of
Ho\v{r}ava's proposal and argue that this version can be understood
as a certain limit of the ghost condensate model. The theory still
problematic since the additional field generically forms caustics
and, again, has a very low strong coupling scale. Also they clarify
some subtleties that arise in the application of the St$¨$uckelberg
formalism to Ho\v{r}ava's model due to it's non-relativistic nature.


 One of the most important topological objects is the cosmic string,
discussed both in f(R) gravity and scalar field theories by the
author\cite{20,21,22}. Our main purpose of this short paper is the
investigation of the special cylindrically symmetric spacetimes
which describes the cosmic strings. Question that we want to answer
is that" Why and when the specific properties of the cosmic string
defected by the new coupling constants in the new kind of the non
relativistic quantum gravity?". In this work we show that the near
axis limit for a cosmic string in the HL theory has a quantum
mechanical origin. it means that we have a minimum mass scale for
the cosmic strings, which enforced us that we must limited ourselves
only to the region of the space near the location of the string.\\
Also we derived the general force exerted by string to a test
particle and by comparing the results from the orbit motion around
the string and comparing our calculations with the known data, we
present a new order magnitude for the coupling constants in this
model which roles as the lorentz breaking terms in the UV limit.

\section{The metric due to an infinite straight string in GR}
The metric due to an infinite straight cosmic string in vacuum is in
its distributional form, arguably the simplest non-empty solution of
the Einstein field equations. The weak-field version (which is
virtually identical to the full solution) was first derived by
 Vilenkin \cite{47}. The full metric was independently
discovered by  Gott \cite{48} and  Hiscock \cite{49}, who matched a
vacuum exterior solution to a simple interior solution containing
fluid with the equation of state
\begin{eqnarray}\nonumber
T^{t}_{t}=T^{z}_{z}=\epsilon
\end{eqnarray}
( $\epsilon$  a constant) and then let the radius of the interior
solution go to zero.The Gott 's work followed directly from a study
of the gravitational field of point particles in 2+1 dimensions
\cite{50}. A more general class of interior solutions was
subsequently constructed by Linet \cite{51}. The Gott-Hiscock
solution is constructed by first assuming a static,
cylindrically-symmetric line element with the general form:
\begin{eqnarray}\nonumber
ds^2=-e^{2\chi}dt^2+e^{2\psi}(dr^2+dz^2)+e^{2\omega}d\varphi^2
\end{eqnarray}
where $\chi,\psi$ and $\omega$ are functions of "r" alone, and "r"
and $\varphi$  are standard polar coordinates on $\mathfrak{R}^2$ .
The metric is generated by solving the non vacuum Einstein equations
\begin{eqnarray}\nonumber
G^{\mu}_{\nu}=-8 \pi T^{\mu}_{\nu}
\end{eqnarray}
The only constraints are that the $\omega$ should be positive and
that the solution should be regular on the axis $r = 0$, so that
$e^{\omega}\sim r$ for small "r"s . Gott \cite{48}  and Hiscock
\cite{49} both assumed $\epsilon$ to be a constant
$\epsilon_{0}$.The more general situation where $\epsilon$ varies
has been discussed by Linet \cite{51}. The exterior metric is the
solution of the vacuum Einstein equations
\begin{eqnarray}\nonumber
G^{\mu}_{\nu}=0
\end{eqnarray}
It was shown that the most general statics, cylindrical-symmetric
vacuum line element is which was first discovered by\emph{ Tullio
Levi-Civit`a }\cite{59},
\begin{eqnarray}\nonumber
ds^2=-r^{2m}c^2dt^2+r^{2m(m-1)}b^2(dz^2+dr^2)+r^{2(1-m)}a_{0}^2d\varphi^2
\end{eqnarray}

It is always possible to set $b$ and $c$ to 1 by suitable rescaling
$t$, $z$ and $r$ , but for present purposes it is more convenient to
retain them as arbitrary integration constants. The interior and
exterior solutions can be matched at any nominated value $r_{0}$ for
$r$ in the interior  solution. It was shown that\cite{42}
\begin{eqnarray}\nonumber
a=1-4\eta
\end{eqnarray}
by noting that the total mass per unit length $\eta$ on each surface
for constant $t$ and $z$ in the interior solution, it is possible to
endow the interior solution with an equation of state more general
than that considered by Gott, Hiscock and Linet \cite{48,49,51}
while preserving the form  of the exterior metric. The mass per unit
length in the interior solution is then typical not equal to the
metric parameter $\frac{1}{4}(1-a)$.

\section{Review of Ho$\check{\textbf{r}}$ava-Lifshitz gravity with detailed balance condition }

 Following from the ADM decomposition of the metric  \cite{31}
, and the Einstein equations, the fundamental objects of interest
are the fields $N(t,x),N_{i}(t,x),g_{ij}(t,x)$ corresponding to the
\emph{lapse }, \emph{shift} and \emph{spatial metric} of the ADM
decomposition.
 In the $(3 + 1)$-dimensional ADM formalism, where the
metric can be written as
 \begin{eqnarray}
ds^2=-N^2dt^2+g_{ij}(dx^{i}+N^{i}dt)(dx^{j}+N^{j}dt)
\end{eqnarray}
and for a spacelike hyper surface with a fixed time, its extrinsic
curvature $K_{ij}$ is

\begin{eqnarray}\nonumber
K_{ij}=\frac{1}{2N}(\dot{g_{ij}}-\nabla_{i}N_{j}-\nabla_{j}N_{i})
\end{eqnarray}
where a dot denotes a derivative with respect to "t" and covariant
derivatives defined with respect to the spatial metric $g_{ij}$ ,
the action of Ho$\check{\textbf{r}}$ava-Lifshitz theory  for $z=3$
is
\begin{eqnarray}
S=\int_{M} dtd^{3}x\sqrt{g} N(\mathcal{L}_{K} - \mathcal{L}_{V} )
\end{eqnarray}
we define the space-covariant derivative on a covector $v_{i}$ as
$\nabla_{i}v_{j}\equiv \partial_{i}v_{j}-\Gamma_{ij}^{l}v_{l}$ where
$\Gamma_{ij}^{l}$ is the spatial Christoffel symbol . The  $g$ is
the determinant of the 3-metric and $N = N(t)$ is a dimensionless
homogeneous gauge field. The kinetic term is
\begin{eqnarray}\nonumber
\mathcal{L}_{K}=\frac{2}{\kappa^2}\mathcal{O}_{K}=\frac{2}{\kappa^2}(K_{ij}K^{ij}-\lambda
K^2)
\end{eqnarray}
Here $N_{i} $ is a gauge field with scaling dimension $[N_{i}] = z -
1$.\\
The \emph{'potential'} term $\mathcal{L}_{V}$ of the
$(3+1)$-dimensional theory is determined by the \emph{principle of
detailed balance }\cite{horava3}, requiring $\mathcal{L}_{V}$ to
follow, in a precise way, from the gradient flow generated by a
3-dimensional action $W_{g}$. This principle was applied to
gravitation with the result that the number of possible terms in
$\mathcal{L}_{V} $ are drastically reduced with respect to the broad
choice available in an '\emph{potential} is
\begin{eqnarray}
\mathcal{L}_{V}=\alpha_{6}C_{ij}C^{ij} - \alpha_{5}\epsilon_{l}^{ij}
R_{im}\nabla_{j}R^{ml} + \alpha_{4} [R_{ij}R^{ij}-
\frac{4\lambda-1}{4(3\lambda-1)} R^2] +\alpha_{2}(R - 3\Lambda_{W})
\end{eqnarray}
Where in it $C_{ij}$ is the \emph{Cotton }tensor\cite{horava3} which
is defined as,
\begin{eqnarray}\nonumber
C^{ij}=\epsilon^{kl(i}\nabla_{k}R^{j)}_{l}
\end{eqnarray}
The kinetic term could be rewrite in terms of the \emph{de Witt
metric} as:
\begin{eqnarray}\nonumber
\mathcal{L}_{K}=\frac{2}{\kappa^2}K_{ij}G^{ijkl}K_{kl}
\end{eqnarray}
Where we have introduced the \emph{de Witt metric}
\begin{eqnarray}\nonumber
G^{ijkl}=\frac{1}{2}(g^{ik}g^{jl}+g^{il}g^{jk})-\lambda g^{ij}g^{kl}
\end{eqnarray}
The inverse of this metric is given by
\begin{eqnarray}\nonumber
G_{ijkl}=\frac{1}{2}(g_{ik}g_{jl}+g_{il}g_{jk})-\tilde{\lambda}g_{ij}g_{kl}\\\nonumber
\tilde{\lambda}=\frac{\lambda}{3\lambda-1}
\end{eqnarray}
Inspired by methods which are used in quantum critical systems and
non equilibrium critical phenomena, Ho$\check{\textbf{r}}$ava
restricts the large class of possible potentials using the principle
of detailed balance outlined above. This requires that the potential
(3) takes the form
\begin{eqnarray}\nonumber
\mathcal{L}_{V}=\frac{\kappa^2}{8}E^{ij}G_{ijkl}E^{kl}
\end{eqnarray}
Note that by constructing $E^{ij} $ as a functional derivative it
automatically transverse within the foliation slice,
$\nabla_{i}E^{ij}=0$. The equations of motion were  obtained in
\cite{Kiritsis}.

\section{About the inconsistency of Ho\v{r}ava gravity }
The action in the ADM formalism contains only first order time
derivatives, which allows to circumvent the problems with the ghosts
appearing in covariant higher order gravity theories \cite{65}. The
higher derivative terms naively become irrelevant in the infrared
and Ho\v{r}ava was argued that the theory reduces to GR at large
distances. However as Blas et al showed, the consistency of the
above proposal is far from being clear. The main concern comes from
the fact that the introduction of a preferred foliation explicitly
breaks the gauge group of GR down to the group of space-time
diffeomorphisms preserving this foliation. As already pointed out by
Ho\v{r}ava, this breaking is expected to introduce extra degrees of
freedom compared to GR. The new degrees of freedom can be persisted
down to the infrared and be leaded to various pathologies
(instabilities, strong coupling) that may invalidate the
theory.\footnote{ An illustration of this phenomenon is provided by
theories of massive gravity where special care is needed to make the
additional degrees of freedom well-behaved [66]}.There have been
several controversial claims about the properties of the extra
freedom degrees. In \cite{66} the new mode was identified among  the
perturbations around a static spatially homogeneous background in
the presence of matter. The mode was argued to be strongly coupled
to matter in the limit when the theory is expected to approach GR,
making it hard to believe that a GR limit exists. It is worth noting
that the mode found in \cite{67} is not propagating: it's equation
of motion does not contain time derivatives \cite{67}. Thus it
remains unclear from this analysis whether this mode corresponds to
a real degree of freedom or can be integrated out as unphysical. The
observation that the extra mode is non-propagating was generalized
in \cite{68} to the case of cosmological backgrounds. The
interpretation of this result given in \cite{68} is that actually
the Ho\v{r}ava gravity is free from additional freedom degrees. Also
it was claimed that the strong coupling is alleviated by the
expansion of the Universe. Finally, the non- linear Hamiltonian
analysis performed in \cite{69} shows that the phase space of
Ho\v{r}ava gravity is 5-dimensional. This result is puzzling: a
normal degree of freedom corresponds to a 2-dimensional phase space;
so the result of \cite{69} suggests that the number of degrees of
freedom in Ho\v{r}ava gravity is two and a half. Two of these
freedom degrees are naturally identified with the two helicities of
graviton. But the physical meaning of the extra "\emph{halfmode}" is obscure.\\
Blas et al. showed that Ho\v{r}ava gravity does possess an
additional light scalar mode. For a general background the equation
of motion for this mode contains time derivatives implying that the
mode is propagating. The peculiarity of Ho\v{r}ava gravity is that
the equation for the extra mode is first order in time derivatives.
The solution still corresponds to waves with a background dependent
dispersion relation and is fixed once a single function of spatial
coordinations is determined as the initial condition in the Cauchy
problem. This explains why this mode corresponds to a single
direction in the phase space. Next we address the consistency of the
Ho\v{r}ava proposal by study the infrared properties of the extra
mode. We find that its dynamics exhibits a number of bad features.
First, the mode becomes singular for static or spatial homogeneous
backgrounds. Namely, the mode frequency diverges in that limit. This
explains why this mode has been overlooked in the previous analysis
of perturbations in Ho\v{r}ava gravity . Second, for certain
(background-dependent) values of spatial momentum the mode becomes
unstable. Again, the rate of the instability diverges if one takes
the static / spatially homogeneous limit for the background metric.
Third, they show that at energies above the certain scale the extra
mode is strongly coupled to itself, and not only to matter. Also
they found that the strong coupling scale is background dependent
and goes to zero for \emph{flat / cosmological} backgrounds. Hence,
the model suffers from a much more severe strong coupling problem
than pointed out in , where the dependence of the strong coupling
scale on the background curvature was ignored. Because of the strong
coupling the Ho\v{r}ava model can be trusted only in a narrow window
of very small energies, way below the Planck scale. This implies
that the Ho\v{r}ava model cannot be considered as consistent theory
of quantum gravity.\\
Later    Charmousis et al.\cite{67}   showed that Ho\v{r}ava gravity
suffers from strong coupling problems, with and without detailed
balance, and is therefore unable to reproduce General Relativity in
the infra-red. They considered the perturbative theory about the
vacuum, yielding two important results. The first considered the
role of detailed balance in these models. As the breaking terms go
zero, They find that the linerized gravitational Hamiltonian
constraint vanishes off -shell. \footnote{This means that linerized
theory breaks down in this limit, just as it does for the
Chern-Simons limit of Gauss-Bonnet gravity}. By comparing field
equations to their counterparts in General Relativity, Charmousis et
al showed that the "\emph{emergent}" Planck length actually diverges
in the limit of detailed balance, in contrast to the original claims
of Ho\v{r}ava. This strong coupling behavior means that the theory
with detailed balance does not have aany sort of perturbative
infra-red limit , explaining the results of  Lu et
al.\cite{10}.Indeed, from the point of view of spherical symmetric
solutions one sees that the putative higher order terms in the IR
are just as important as the "lower" order terms. In summary, with
detailed balance, one can never hope to recover GR in the infra-red
for the following reason: General Relativity admits an effective
linearised description beyond the Schwarzschild radius of a source,
but in Ho\v{r}ava gravity with detailed balance, strong coupling
prevents an effective linearized description on any scale.Further
discussions may be found it \cite{Prob}.
\section{Thermodynamics and Dark energy in Horava gravity}
If we assume that the cosmological scenario of a universe governed
by Horava-Lifshitz gravity, it is a natural problem to an
investigation of its thermodynamic properties, and in particular of
the generalized second thermodynamic law. The validity of the
generalized second law of thermodynamics in a universe governed by
Horava-Lifshitz gravity has been discussed by Jamil et al in
\cite{GSL}. They calculated  the entropy time-variation for the
matter fluid and, using the modified entropy relation, that of the
apparent horizon itself and found that under detailed balance the
generalized second law was generally valid for flat and closed
geometry and it was conditionally valid for an open universe, while
beyond detailed balance it was only conditionally valid for all
curvatures. They followed the exact and robust approach, that is
they used the  modified entropy relation as it has been calculated
in the specific context of Horava-Lifshitz gravity. Under the
equilibrium assumption between the universe interior and the
horizon, which is expected to ba valid at late cosmological times,
they found that the generalized second law is only conditionally
valid. The possible violation of the generalized second
thermodynamical law in Horava-Lifshitz cosmology could lead to
various conclusions. Further the  dark energy was discussed by
setare et al \cite{Holographic}. They nicely investigated the
holographic dark energy scenario with a varying gravitational
constant in a flat background in the context of Horrava-Lifshitz
gravity and immediately, extracted and determined the evolution of
the dark energy density parameter. Also they discussed some non
trivial cosmological implications of this holographic model. Also
they  evaluated the dark energy equation of state for low redshifts
even when the model contains a time varying gravitational constant.

\section{Exact solutions}
Considering the non-static, cylindrically symmetric solutions with
the metric ansatz
\begin{eqnarray}
ds^2=-N(t)^2dt^2+\frac{1}{N(t)^2}[dr^2+\Phi(r)^2dz^2+\Psi(r)^2
d\varphi^2]
\end{eqnarray}
For simplicity we decompose the spatial metric as a conformal by
another diagonal simple form:
\begin{eqnarray}\nonumber
g_{ij}dx^idx^j=\frac{1}{N(t)^2}\gamma_{ij}dx^idx^j\\
\gamma_{ij}=diag(1,\Phi(r)^2,\Psi(r)^2)
\end{eqnarray}
This may or may not represent a cosmic string, and it may have
singularities and/or event horizons.

 In synchronous time $t$ , the Cylindrical \emph{ADM } metric
has $N_{i}=0$ , where $\Phi(r),\Psi(r)$ are the usual  Weyl metric
functions. Too in GR and only for vacuum solutions the metric
function $\Psi(r)=r$ and in another cases is determined from a
quadrature on another metric functions $\Phi(r)$\cite{25}. There is
another general choose for metric form but since in this paper our
main goal is to investigate static cosmic strings in analogous for
usual GR samples, we limited ourselves to this simple but applicable
gauge. On this background,
\begin{eqnarray}
K_{ij}=-\frac{\dot{N}}{N^4}\gamma_{ij}\\K=K_{i}^{i}=-3\frac{\dot{N}}{N^2}\\K_{ij}K^{ij}=3(\frac{\dot{N}}{N^2})^2
\end{eqnarray}
Following Sotiriou and et al \cite{23} we use from a general full
classical action ,
\begin{eqnarray}
S=\int[T(K)-V(g)]\sqrt{g}Nd^3xdt
\end{eqnarray}
Where
\begin{eqnarray}
T(K)=g_{K}{(K^{ij}K_{ij}-K^2)+\xi K^2}
\end{eqnarray}
This is the general kinetic term corresponds to the limit
$\xi\rightarrow 0$. Since the kinetic action is (by definition)
chosen to be dimensionless, we have set the critical exponent $z=3$
to make $g_{K}$ dimensionless, then provided $g_{K}$ is positive one
can without loss of generality rescale the time and /or space
coordinates to set $g_{K}\rightarrow 1$. Now consider the following
form for potential term:
\begin{eqnarray}
V(g)=g_{0}\zeta^6+g_{1}\zeta^4R+g_{2}\zeta^2R^2+g_{3}\zeta^2R_{ij}R^{ij}+\\\nonumber
g_{4}R^3+g_{5}R(R_{ij}R^{ij})+g_{6}R_{j}^{i}R_{k}^{j}R_{i}^{k}+g_{7}R\nabla^2R+g_{8}\nabla_{i}R_{jk}\nabla^{i}R^{jk}
\end{eqnarray}
As in \cite{24} was stated, suitable factors of $\zeta$ are
introduced to ensure the coupling $g_{a}$ are all dimensionless.
Without loss of generality we can rescale the time and space
coordinations to set both of the $g_{K}\rightarrow1$ and
$g_{1}\rightarrow-1$.From normalization of the Einstein$-$Hilbert
term, we see that in physical units $c\rightarrow1$
\begin{eqnarray}\nonumber
(16\pi G_{Newton})^{-1}=\zeta^2\\\nonumber
\Lambda=\frac{g_{0}\zeta^2}{2}
\end{eqnarray}
so that $\zeta$ is identified as the Planck scale. The cosmological
constant is determined by the free parameter $g_{0}$, and obviously
$g_{0} \sim 10^{-123}$. In particular, the way Sotiriou had set this
up , now we are free to choose the Newton constant and cosmological
constant independently (and so to be compatible with observation).
In contrast, in the original model presented in \cite{horava3}, a
non-zero Newton constant requires a non-zero cosmological constant,
of the wrong sign to be compatible with
cosmological observations\cite{Kiritsis,30}.\\
  For a special choose of our
metric (4) the Ricci scalar and non-vanishing components of Ricci
Tensor are :
\begin{eqnarray}
R_{11}=\frac{\Phi''}{\Phi}+\frac{\Psi''}{\Psi}\\\nonumber
R_{22}=\Phi^2(\frac{\Phi''}{\Phi}+\frac{\Phi'}{\Phi}\frac{\Psi'}{\Psi})\\\nonumber
R_{33}=\Psi^2(\frac{\Psi''}{\Psi}+\frac{\Phi'}{\Phi}\frac{\Psi'}{\Psi})\\R=2N(t)^2(\frac{\Phi''}{\Phi}+
\frac{\Psi''}{\Psi}+\frac{\Phi'}{\Phi}\frac{\Psi'}{\Psi})
\end{eqnarray}
Here, a prime denotes a derivative with respect to r.By substituting
(6,7,8) in (10) and (12,13) in (11) and all in (9) we obtain  the
following form of action:
\begin{eqnarray}
S=\int dt
d^3xN^3\Phi\Psi[3(3\xi-2)(\frac{\dot{N}}{N^2})^2-f(R)-(g_{3}\zeta^2+g_{5}R)H-g_{6}W-g_{7}B-g_{8}Y]\equiv\int
dt d^3x \Xi
\end{eqnarray}
Where in it,
\begin{eqnarray}\\\nonumber
f(R)=g_{0}\zeta^6-\zeta^4R+g_{2}\zeta^2R^2+g_{4}R^3
\\\nonumber
H=R_{ij}R^{ij}=N^4[(\frac{\Phi''}{\Phi}+\frac{\Psi''}{\Psi})^2+(\frac{\Phi''}{\Phi}+\frac{\Phi'}{\Phi}\frac{\Psi'}{\Psi})^2
+(\frac{\Psi''}{\Psi}+\frac{\Phi'}{\Phi}\frac{\Psi'}{\Psi})^2]\\\nonumber
W=R_{j}^{i}R_{k}^{j}R_{i}^{k}=N^6[(\frac{\Phi''}{\Phi}+\frac{\Psi''}{\Psi})^3+(\frac{\Phi''}{\Phi}+\frac{\Phi'}{\Phi}\frac{\Psi'}{\Psi})^3
+(\frac{\Psi''}{\Psi}+\frac{\Phi'}{\Phi}\frac{\Psi'}{\Psi})^3]\\\nonumber
B=R\nabla^2R=4N^4(\frac{\Phi''}{\Phi}+\frac{\Psi''}{\Psi}+\frac{\Phi'}{\Phi}\frac{\Psi'}{\Psi})(\frac{\Phi'}{\Phi}R'+\frac{\Psi'}{\Psi}R'
+R'')\\\nonumber
Y=\nabla_{i}R_{jk}\nabla^{i}R^{jk}=N^6[((\frac{\Phi''}{\Phi}+\frac{\Psi''}{\Psi})')^2+
[(\Phi^2(\frac{\Phi''}{\Phi}+\frac{\Phi'}{\Phi}\frac{\Psi'}{\Psi}))'(\frac{1}{\Phi^2}
(\frac{\Phi''}{\Phi}+\frac{\Phi'}{\Phi}\frac{\Psi'}{\Psi}))']
+[\Phi\rightarrow\Psi]]
\end{eqnarray} The extreme functions are the solutions of the
\textrm{Euler-Lagrange} equations that are obtained by setting the
all variational derivatives of the functional with respect to each
function $X\equiv(\Phi(r),\Psi(r),N(t)) $equal to zero. The Ritz
variational principle affords a powerful technique for the
approximate solution of (9).The result is an upper bound on the
corresponding eigenvalue and optimal values for the parameters of
(9) \footnote{Variational Bound}. Variational Bound can also be used
to extremize  general functional given appropriate trial functions.
We remind that if we consider a system with the Lagrangian with
linear terms of curvature $R$ we must recover the GR solutions i.e,
a non static cylindrical solution with cosmological constant.
However It is proper , if we make an analytic continuation of
Coordinations $r,t$ , namely, we obtain the Tian solution \cite{26}
in a special coordinations or a non static solution which is
\emph{pure radiation field} generated from a flat space-time  and
has a Weyl tensor of type N .The metric of this space-time is
described by the \emph{Rao} line element\cite{27},
\begin{eqnarray}
 ds^2=e^{k(t-r)}(dr^2-dt^2)+r^2d\varphi^2+dz^2
\end{eqnarray}
This is a special case of non static Weyl gauges which we used in
witting (4). For this metrics with null vector fields it was shown
that \cite{28} in our notations $\Psi(r)=r$. The equations of motion
due to the variation of metric functions are more complicated and we
do not present them here. As a simple but physically important case
we seeking only those solutions which can be described the cosmic
strings . We choose a very restricted gauge as,
\begin{eqnarray}
 \Phi(r)=1,N(t)=\emph{const}
\end{eqnarray}
In GR these constraints leads to a static\emph{ cosmic strings} and
also in \emph{metric f(R) gravity}\cite{21,22}. The resulting metric
with new parameters $(\zeta,g_{0}...g_{8})$ may be so interesting.
In order to obtain the solution, let us substitute the metric ansatz
(4) with constraints (17)  into the action, and then vary the
function $\Psi(r)$.The same process was done in spherical symmetry
in \cite{16} This is possible because the metric ansatz shows all
the allowed singlets which are compatible  with the $SO(3)$ action
on the $S^2\times \mathbb{R}$. The resulting reduced Lagrangian, up
to an overall scaling constant, is given by
\begin{eqnarray}
 \mathfrak{L_{0}}=\Psi(r)[-f_{0}-(g_{3}\zeta^2+g_{5}R_{0})H_{0}-g_{6}W_{0}-g_{7}B_{0}-g_{8}Y_{0}]
\end{eqnarray}
Where
\begin{eqnarray}\nonumber
 R_{0}=2\frac{\Psi''}{\Psi}\\\nonumber
 f_{0}=f(R_{0})\\\nonumber H_{0}=\frac{1}{2} R_{0}^2\\\nonumber W_{0}=\frac{1}{4}
 R_{0}^3\\\nonumber B_{0}=2 R_{0} (R_{0}'\frac{\Psi'}{\Psi} +R_{0}'')\\\nonumber
 Y_{0}=R_{0}'^2+(\Psi^2R_{0})'(\frac{R_{0}}{\Psi^2})'
\end{eqnarray}
 The functional (14) with reduced Lagrangian (18) is in the form,
\begin{eqnarray}
S=(t_{2}-t_{1})\int  d^3\Pi(\Psi,\Psi',...,\Psi^{6})
\end{eqnarray}
Where
\begin{eqnarray}\nonumber
\Psi^{a}=\frac{d^a\Psi}{dr^a}
\end{eqnarray}
By using a general variational principle applied to this higher
order function[25]we can write all equations of motion for metric
function $\Psi$,
\begin{eqnarray}
 \frac{d \mathfrak{L_{0}}}{d
 \Psi}+\sum^6_{a=1}(-1)^a\frac{d^a}{dr^a}[\frac{\partial\mathfrak{L_{0}}}{\partial\Psi^{a}}]=0
\end{eqnarray}
Because the equation of motion contain up to eight derivative terms,
it is difficult to find the exact solutions. In order to understand
the behavior of  solutions in Ho$\check{\textbf{r}}$ava gravity we
try to solve the equations of motion in a special case. By our
inspirations from GR we know that a cosmic string has a linear
functionality as $\Psi=ar+b$. The constant  $b$ may be turned to
zero by changing scales along the $t$ and $z$ axes and choosing the
zero point of the $r$ coordinate.Since our model essentially is not
GR, we expected that an adhoc assumption for metric function as
\begin{eqnarray}
 \Psi=(Ar+B)^m
\end{eqnarray}
For a general value of $A,m,B$  the solution to the equations of
motion is consistent only with the following  cases. Considering $m$
as a real parameter, one must note that for
$m=1,B=0,A^2=1-4\eta$($\eta$ mass per length of string) we have a
cosmic string. By considering this constraint we find the field
equation (20) with solution (21) is satisfied identically.
\subsection*{a: Solutions with $g_{8}=0,g_{7}=0,B=0$}
In this case by substituting (21) in field equation (20) we obtain
\begin{eqnarray}\nonumber
m_{i}=-3,4,i=1,2
\end{eqnarray}
Consequently we can set $B=0$ in metric function (21).Thus the most
general solution for (4) is one of the two possible functions
\begin{eqnarray}
\Psi(r)=(Ar)^{m_{i}}
\end{eqnarray}
Thus,
\begin{eqnarray}
ds^2=-dt^2+dr^2+dz^2+(Ar)^{2m_{i}}d\varphi^2
\end{eqnarray}
In this case there is an arbitrarily in choosing another
\emph{coupling coefficients} $(g_{2}...g_{6})$.The Ricci scalar is
given by
\begin{eqnarray}
R=2\,{\frac {m_{i} \left( m_{i}-1 \right) }{{r}^{2}}}
\end{eqnarray}
The solution has a curvature singularity at $r = 0$ for general
$m_{i}\neq 0,1$. The only non zero component of the Riemann Tensor
is:
\begin{eqnarray}\nonumber
 R_{rzrz}={\frac { \left( Ar \right) ^{2\,m}m \left(
m-1 \right) }{{r}^{2}}}
\end{eqnarray}
 The a singularity structure of the solution (23) is apparent from
its Kretschmann scalar:
\begin{eqnarray}\nonumber
 \mathfrak{R}={\frac { \left( Ar \right) ^{4\,m}m^2 \left(
m-1 \right)^2 }{{r}^{4}}}
\end{eqnarray}
then the Kretschmann scalar has a naked singularity at $r = 0$. In
GR, the cosmological horizon(s) for cylindrically symmetric
space{times is discussed in detail by Wang \cite{Singularity}.

\subsection*{b: Solutions with $g_{8}=0,g_{7}=0,g_{2}=0, B\neq0$}
In this case as the previous section, substituting a general form of
(21) in (20) leads to the similar values for $m$ . It is particular
interest to investigate the $m=-3$  solution, in which case, the
\emph{coupling coefficients} $(g_{3}...g_{6})$ constrained to
$g_{6}=2g_{5}+4g_{4}$. This is one of the good results of this
paper.We obtained a restriction on some constants of the model.In
this case the general solution can be written as
\begin{eqnarray}
ds^2=-dt^2+dr^2+dz^2+(Ar+B)^{2m_{i}}d\varphi^2
\end{eqnarray}
The solution has a curvature singularity at $r = \frac{-B}{A}$ for
general $B<0$.We mention here that the cases $m=0,1$ is satisfied
field equations without any limitation. Specially the case $m=1$ is
so interesting. Since it represents the usual familiar line element
of a static cosmic string. In this case there is an arbitrarily in
choosing another \emph{coupling coefficients} $(g_{2}...g_{6})$.The
Ricci scalar is given by
\begin{eqnarray}
R=2A^2\frac {m_{i} ( m_{i}-1 ) }{(Ar+B)^{2}}
\end{eqnarray}
The solution has a curvature singularity at $r =\frac{-B}{A}$ for
general $m_{i}\neq 0,1$. The only non zero component of the Riemann
Tensor is:
\begin{eqnarray}\nonumber
 R_{rzrz}=m \left( m-1 \right)  \left( Ar+B \right) ^{2\,m-2}{A}^{2}
\end{eqnarray}
 The a singularity structure of the solution (25) is apparent from
its Kretschmann scalar:
\begin{eqnarray}\nonumber
 \mathfrak{R}=(m \left( m-1 \right))^2  \left( Ar+B \right) ^{4m-4}{A}^{4}
\end{eqnarray}
then the Kretschmann scalar has a naked singularity at $r =
\frac{-B}{A}$.

\subsection*{c:Solutions with $g_{8}=0,g_{7}=0,g_{4}+\frac{1}{2}g_{5}+\frac{1}{4}g_{6}=0,g_{2}+\frac{1}{2}g_{3}=0$}
The Lagrangian (18) in this case has only three independent coupling
constants and reduces to the following form:
\begin{eqnarray}
 \mathfrak{L_{0}}=\zeta^4(-g_{0}\Psi+2\zeta^2\Psi'')
\end{eqnarray}
the field equation (20) will be
\begin{eqnarray}
g_{0}=0
\end{eqnarray}
This term has a significant physical meaning. If we refer to the
previous relations between parameters of model we observe that if
this condition holds, the cosmological constant must be vanished.
Thus this model describes a non classical (for appearance a second
order derivatives of matter field in action) system with no
potential term. Indded the action (19) may be integrated to obtain:
\begin{eqnarray}
S=4\zeta^6\pi(t_{2}-t_{1})l\int  dr\Psi''r^2
\end{eqnarray}
Where in it we assumed that the cylindrical coordinations $z$ is
bounded in interval $(0,l)$.If we carry out a part by part
integration on the radial part of this integral and by assumption
that our Potential function $\psi$ may be bounded and posses
suitable boundary conditions(a well posses function) entirely away
from the origin of radial coordinations $r=0$ to infinity, Note that
the action (28) can be written as
\begin{eqnarray}\nonumber
S=8\zeta^6\pi(t_{2}-t_{1})l\int  dr\Psi r+ B
\end{eqnarray}
where B is a surface term, which must be chosen so that the action
has an extreme under variations of the fields with appropriate
boundary conditions. One demands that the fields approach the
classical solutions at infinity. Varying the action ,we find the
boundary term
\begin{eqnarray}\nonumber
\delta B=-(t_{2}-t_{1})N_{0}\delta M
\end{eqnarray}
The boundary term B is the conserved charge associated to the
\emph{improper gauge transformations} produced by time
evolution.Here $M$ and  $N_{0} $  are conjugate pairs. Therefore
when one varies $M$, $N_{0}$ must be fixed. Thus the boundary term
should be in the form
\begin{eqnarray}\nonumber
 B=-(t_{2}-t_{1})N_{0} M+B_{0}
\end{eqnarray}
where $B_{0}$  is an arbitrary constant, which should be fixed by
some physical considerations; for example, in \emph{topological
black hole} case with arbitrary constant scalar curvature horizon.
Mass vanishes when  black hole's horizon goes to zero\cite{19}. For
details, see\cite{40}. We will not go further in detail. The
dynamics of the metric function in this case is not determined
without more mathematical features of variational calculus which is
found in any textbook in this field as which is discussed in
\cite{24}.
\section{The real cosmic string}
Now let we impose the next constraints,
\begin{eqnarray}
g_{8}=0,g_{7}=0,g_{4}+\frac{1}{2}g_{5}+\frac{1}{4}g_{6}=0,g_{2}+\frac{1}{2}g_{3}\neq0
\end{eqnarray}
From the action, we can obtain the equation of motion as
\begin{eqnarray}
-g_{0}\zeta^6+2g_{9}R-6g_{9}\frac{d^2R^2}{dr^2}=0
\end{eqnarray}
where $R=2\frac{\Psi''}{\Psi}$ is the Ricci scalar. The function
$R(r)$ can be obtained as
\begin{eqnarray}
R(r)=\beta+\frac{1}{C^2}sn(Ar,D)^2
\end{eqnarray}
Where the \emph{Jacobi elliptic functions sn} is in turn defined in
terms of the amplitude function \emph{JacobiAM}\cite{39}

\begin{eqnarray}\nonumber
sn(z,k)\equiv JacobiSN(z,k) = sin(JacobiAM(z,k))
\end{eqnarray}
and

\begin{eqnarray}\nonumber
\phi =
JacobiAM(\int^{\phi}_{0}\frac{d\phi}{(1-k^2\sin(\theta)^2)^{1/2}},k),
\phi\in (-3/2,3/2)
\end{eqnarray}

Where in (31)

\begin{eqnarray}
C=\xi e^{i\theta}=0.19+0.73 i\\D=\eta
e^{i\rho}=0.86+0.50i\\A=\pm\sqrt{\frac{\beta}{15}}\frac{1}{2\xi}e^{i\theta}\\
\beta^3=\frac{15}{2}\alpha\\\alpha=\frac{g_{0}\zeta^6}{6g_{9}},g_{9}=g_{4}+\frac{1}{2}g_{5}+\frac{1}{4}g_{6}
\end{eqnarray}

As in [23]is proved that the $\zeta=M_{pl}$,
$g_{0}=\frac{2\Lambda}{M_{pl}^2}$.for metric(4) with (17) the Ricci
scalar is

\begin{eqnarray}
R(r)=2\frac{\Psi''}{\Psi}
\end{eqnarray}

The general solution for (37) with (31) is so complicated. Instead
of doing that, we focused ourselves only to the near axis $r\approx
0$ behavior of (31). It is adequate to define a very important
physical radial scale

\begin{eqnarray}
r_{0}=\frac{5.670}{M_{pl}}(\frac{g_{9}}{g_{0}})^{\frac{1}{6}}
\end{eqnarray}

With this length scale the meaning of \emph{near axis } limit is
thinkable as the following expression

\begin{eqnarray}
r<<r_{0}
\end{eqnarray}

Not that in this limit we do not tend to the origin. In this good
reasonable physical approximation the differential equation (37)
(albeit after expansion by series (31) in terms of
$\frac{r}{r_{0}}$) can be solved easily. The solution is written in
terms of \emph{Whittaker} functions $M,W$\cite{39},

\begin{eqnarray}
\Psi(r)=\frac{1}{\sqrt{r}}[(c_{1}WhittakerM(-\frac{\beta
r_{0}}{4}\sqrt{2},\frac{1}{4},\sqrt{2}\frac{r^2}{r_{0}})+c_{2}WhittakerW(-\frac{\beta
r_{0}}{4}\sqrt{2},\frac{1}{4},\sqrt{2}\frac{r^2}{r_{0}})]
\end{eqnarray}

once again we impose the near axis limit on (40).The result up to
order one is

simple\footnote{$\Gamma(s)=\int_{0}^{\infty}e^{-t}t^{s-1}dt$}

\begin{eqnarray}
\Psi(r)=ar+b\\
a\equiv
(\frac{\sqrt{2}}{r_{0}})^{3/4}(c_{1}-2\frac{\sqrt{\pi}c_{2}}{\Gamma(\frac{1}{4}+\beta
r_{0}\frac{\sqrt{2}}{4})})\\
b\equiv\frac{\sqrt{\pi}(\frac{\sqrt{2}}{r_{0}})^{1/4}c_{2}}{\Gamma(\frac{3}{4}+\beta
r_{0}\frac{\sqrt{2}}{4})}
\end{eqnarray}

Again, $c_{1},c_{2}$  are  integration constants and $c_{2} $ could
be set to zero. Similar to the case of GR, we find the metric of a
\emph{cosmic string}
\begin{eqnarray}
ds^2=-dt^2+dr^2+dz^2+(ar)^2d\varphi^2
\end{eqnarray}
The metric (44) is locally flat and can be brought to a
\emph{Minkowski}  form in any region not surrounding the string.
This implies that the presence of the string has no effect on
physical process in such a region. In particular, a test particle
which is initially at rest relative to the string will remain at
rest and will not experience any gravitational force. Although the
space around the string is locally fiat its global structure is
different from that of Euclidean space.

The parameter $a$ in (44) can be expressed in terms of mass per unit
length of spacetime $\eta$\cite{41,42},
\begin{eqnarray}
a=1-4\eta
\end{eqnarray}

The mass per unit length of spacetime $\eta$ is found to be
\begin{eqnarray}
\eta=0.25-0.54109c_{1}(M_{pl})^{\frac{1}{2}}(\frac{\Lambda}{g_{4}+\frac{1}{2}g_{5}+\frac{1}{4}g_{6}})^{\frac{1}{8}}
\end{eqnarray}
Clearly $\eta<0.25$ is for all positive values of $c_{1}$. Now we
compare this function with another one which was found by Aryal et
al \cite{58} . They constructed a solution which describes two
Schwarzschild black holes held apart by a system of cosmic strings,
by generalizing a solution, due to Bach and Weyl. Their metric  is
vacuum at all points away from the  axis $r=0$ and describes two
black holes with masses $m_{1}$ and $m_{2}$ located on the  $z$ axis
, and so separated by a z-coordination distance $2d$. The black
holes are held in place by conical singularities along the different
axial segments. The  effective mass per unit length $\eta$ of any of
the conical segments depends only on the limitation of the value of
the metric function  on the axis. For all positive values of
$m_{1}$,$m_{2}$ and $d$, In the original Bach-Weyl solution the mass
per unit length $\eta_{ext}$ of the exterior segments was assumed to
be zero and This forces $\eta_{int}<0$ and the interior Bach-Weyl
segment are normally characterized as a 'strut' rather than a
string. However in (46), all
 segments will have non-negative masses per unit length if
\begin{eqnarray}\nonumber
 c_{1}>0.46203
(M_{pl})^{-\frac{1}{2}}(\frac{\Lambda}{g_{4}+\frac{1}{2}g_{5}+\frac{1}{4}g_{6}})^{-\frac{1}{8}}
\end{eqnarray}
In the particular case
\begin{eqnarray}\nonumber
 c_{1}=0.46203
(M_{pl})^{-\frac{1}{2}}(\frac{\Lambda}{g_{4}+\frac{1}{2}g_{5}+\frac{1}{4}g_{6}})^{-\frac{1}{8}}
\end{eqnarray}

the interior segment vanishes.

It should be also noted that the parameter $\eta$ appearing in this
derivation plays two essential independent roles:\\
 \emph{one as a
measurement of the strength of the gravitational field in the
exterior metric}  and a second one \emph{as the integrated mass per
unit length of the interior solution} .In GR it is possible to endow
the interior solution with an equation of state more general than
that considered by Gott, Hiscock and Linet \cite{48,49,51} while we
preserve the form  of the exterior metric, The mass per unit length
in the interior solution is not typically equal to the metric
parameter $\frac{1}{4}(1-a)$. For this reason the symbol $\eta$
reflects a geometric property of the exterior metric which only will
be referred as the \emph{gravitational mass per unit length} of the
spacetime.

The metric (44) with (45,46) describes Minkowski spacetime with a
wedge of angular extent that $\Delta\varphi=8\pi\eta$ has removed
from each of the constant surfaces $t$ and $z$. The apex of each
wedge lies on the axial plane $r = 0$, and the sides of the wedge
are \emph{glued} together by forming what is sometimes referred as
conical spacetime .The fact that the metric (44) is locally
Minkowskian implies that the Riemann tensor is zero everywhere
outside the axial plane, and therefore when a test particle moving
through the metric would experience no tidal forces. In particular,
such a particle would not be accelerated towards the string.
Therefor a local observer should be unable to distinguish a
preferred velocity in the z-direction; where any gravitational force
in the radial direction would destroy this symmetry. When it is
combined with the other symmetries of the metric, this property
forbids gravitational acceleration in any direction. Incidentally,
Mark Hindmarsh and Andrew Wray \cite{52} shown, by detailed analysis
of the geodesics in a general Levi-Civita spacetime , that
\emph{gravitational lensing} with a well-defined angular separation
between the images is possibly only in the specialized string case
$m = 0$. When $\Lambda=0$, $\eta$ goes back to $0.25$, the effect of
higher derivative terms disappears. As one want, General Relativity
is not recovered because the extra freedom degrees which are
presented in the full theory all are not decouple. On the contrary,
one of those freedom degrees becomes strongly coupled, and one
recovers General Relativity  with an additional strongly coupled
scalar. It is difficult to see how this would correspond to a better
choice since we move away from testable regions of GR. Therefore,
evidence of this mass function absent in classical local tests of
general relativity which it's implement may be weak and moving
slowly sources.In Ho$\check{\textbf{r}}$ava gravity we have seen
that we have no reliable linearized theory to work with due to
strong coupling of an extra scalar degree of freedom. Even if it was
tractable, it seems unlikely that a non-linear analysis could
recover the successes of the General Relativity for cosmic strings
in this case.One can then easily be seen that three of the coupling
constants $g_{i},i=4,5,6$ cannot be set to zero. Thus, in general,
there are cases where what appears should be violation of a symmetry
is just a  new choice of mass function for cosmic string. The same
happens with Ho$\check{\textbf{r}}$ava theory. It looks that it
violates four-dimensional covariance but this is just because it is
written in a specific gauge, specified by the ADM frame, which can
be used just because of the four-dimensional covariance of the
theory (and the corresponding constraints).
\section{About  the existence of non-static cylindrical solutions}
The cylindrical symmetric strings are not the single class of cosmic
strings. As was shown by several authors ,in GR there are both local
and global non static cosmic strings in the context of Lyra geometry
\cite{Lyra}, static and non-static plane symmetric cosmic strings in
Lyra manifold \cite{REDDY}, non static self gravitating fluids
\cite{Roy} and non static cylindrical vacuum solutions
\cite{radhakrishna}. Lyra \cite{Lyra} proposed a modification of
Riemannian geometry by introducing a gauge function in to the
structure less manifold, as a result of which the cosmological
constant arises naturally from the geometry. Several authors have
studied cosmological models based on this manifold with a constant
gauge vector in the time direction. Non-static plane symmetric
cosmic string model is quite similar to the non-static plane
symmetric Zeldovich model $p=\rho$ obtained by Reddy and Innaiah
\cite{Reddy-Innaiah} and Reddy \cite{Reddy88} in general relativity.
This  model reduces to empty space-time discussed by Bera
\cite{Bera} in general relativity. In Lyra geometry there is  a
global string , the energy momentum tensor components are calculated
from the action density for a complex scalar field $\psi$ along with
a typical  potential. But finding these classes of solutions in
Horava gravity are most complicated and we can not present them
here. But as  good problems for further considerations we can treat
them in future plan.

 \section{The self-force on a massive particle near a Ho$\check{\textbf{r}}$ava-Lifshitz straight string}

Observational constraints in HL theory have been discussed by
several
 authors\cite{Obs}.In the last sections we examine some possible
 constraints on the parameters by calculating the self force in the
 field of a cosmic string which was obtained in the previous
 section.
In GR we know that a charged particle at rest in the spacetime
experiences a repulsive self-force \cite{53}, while fluctuations of
the quantum vacuum near a straight string have a non-zero
stress-energy tensor and can induce a range of interesting effects
\cite{54,55,56}. In the weak-field approximation, the gravitational
field due to a particle of mass "m" at rest at a distance "a" from a
straight string is the most convenient which is calculated by
transforming to the \emph{Minkowski} form (44) of the metric and for
fixing the coordinations so that the particle lies at $z = 0$ and
$\varphi=\varphi_{0}=\pi(1-4\eta)$.Now we want to generate a
meaningful expression for the self-force on the particle. The
formula for the gravitational self-force was first derived by
\emph{Dmitri Gal'tsov} in  \cite{57}, although the electrostatic
case, which is formally identical, was analyzed by Bernard Linet
four years earlier \cite{53}. by following \emph{Gal'tsov}, it is
instructive to write the self-force in the form
\begin{eqnarray}
\overrightarrow{F}=-\frac{G m^2 \eta}{a^2}f(\eta)\hat{r}
\end{eqnarray}
where
\begin{eqnarray}
f(\eta)=\frac{1}{4\pi\eta}\int^{\infty}_{0}[\frac{\sinh(\pi
u/\varphi_{0})\pi/\varphi_{0}}{\cosh(\pi
u/\varphi_{0})-1}-\frac{\sinh(u)}{\cosh(u)-1}]\frac{d
u}{\sinh(\frac{u}{2})}
\end{eqnarray}

\section{The value of the coupling constants   $g_{4},g_{5},g_{6}$ obtained  from analysis of bound circular
orbits}

 In GR, the fact that the self-force $\overrightarrow{F}$ is
central has given rise to the common misapprehension that bound
circular orbits exist for massive particles in the neighborhood of a
straight cosmic string. It is true that if (47)  should be continue
to hold for a moving particle, then circular orbits would exist with
the standard Newtonian dependence of the orbital speed
\begin{eqnarray}
v_{circ}=\sqrt{\frac{Gm\eta f(\eta)}{a}}
\end{eqnarray}
Thus, for example\cite{42}, a body with $m = 7 \times 10^{22}$ kg
(roughly equal to the mass of the Moon) could orbit a GUT string
with $\eta = 10^{-6}$ at a distance $a = 4 \times 10^8$ m (the mean
Earth-Moon distance) if $v_{circ}\approx 0.1 m s^{-1}$ , which is
about $1/10 000$ th of the Moon's actual orbital speed around the
Earth. Substituting this approximated value of $\eta$ in (46) we
obtain
\begin{eqnarray}
g_{9}\approx 481.55927 (c_{1}^2M_{pl})^4 \Lambda
\end{eqnarray}
Remember that $c_{1}$ is fixed by invoke quantum theory of gravity.
The relation (50) is fundamentally important. It is related between
quantum gravity and the cosmological observations.

 \section{Conclusion}
In conclusion, we found cylindrical symmetric solutions with
arbitrary scalar curvature  in Ho$\check{\textbf{r}}$ava-Lifshitz
theory,by generalizing the static cylindrical symmetric solutions in
GR ,.We found that there exists solutions only in special choose of
the coupling constants .One of the solutions has a near axis
behavior as cosmic string. For this solution we can define a finite
mass per unit length spacetime .Such an explicit term occurs in the
occasion of considering quantum corrections to cosmic string line
element.In our mass per unit length expression, there is a
undetermined constant $c_{1}$. To fix the constant $c_{1}$, one has
to invoke quantum theory of gravity. The self-force on a massive
particle near a Ho$\check{\textbf{r}}$ava-Lifshitz straight string
is re calculated.By analyzing  bound circular orbits we derived a
new value for the coupling constants of Ho$\check{\textbf{r}}$ava
theory which seems that there is a new observational method for
estimating the validity of the Ho$\check{\textbf{r}}$ava model in
the context of cosmology.

\section{Acknowledgement}
D.Momeni thanks the editor of IJTP, Professor \emph{Heinrich Saller}
and the anonymous referees made excellent observations and
suggestions which resulted in substantial improvements of the
presentation and the results. Also the author thanks  Anzhong Wang ,
Jared Greenwald (Baylor-USA),Antonios Papazoglou(University of
Portsmouth),Sourish Dutta(Vanderbilt University), Domenico
Orlando(IPMU-Japan),Lorenzo Iorio(Istituto Nazionale di Fisica
Nucleare (INFN)-Italy), Miao Li , Yi Pang(ITP-China), M.R.
Setare(Department of Science,Bijar, Iran),E.N. Saridakis(University
of Athens), and Nobuyoshi Ohta(Kinki University-Japan) for useful
comments and valuable suggestions.The author would like to thank
Shahram Khosravi  (Tarbiat Moa'llem University, Tehran) for kind
hospitality and support. This work is supported by Tarbiat Moa'llem
University, Tehran.

\end{document}